\documentclass[twocolumn,showpacs,preprintnumbers,superscriptaddress]{revtex4}
\usepackage{amsmath}
\usepackage{amssymb}
\usepackage{amsmath}
\usepackage{epsfig}
\usepackage{graphicx}
\usepackage{inputenc}
\inputencoding{latin1}
\usepackage{ulem}

\def\slashchar#1{\setbox0=\hbox{$#1$}
   \dimen0=\wd0 \setbox1=\hbox{/} \dimen1=\wd1
   \ifdim\dimen0\big>\dimen1 \rlap{\hbox to \dimen0{\hfil/\hfil}} #1
   \else  \rlap{\hbox to \dimen1{\hfil$#1$\hfil}} / \fi}

\newcommand{\ud}{\mathrm{d}}
\newcommand{\be}{\begin{equation}}
\newcommand{\ee}{\end{equation}}
\newcommand{\bea}{\begin{eqnarray}}
\newcommand{\eea}{\end{eqnarray}}

\newcommand{\Appendix}[1]%
    {%
     \section{#1}%
      }

\begin{document}

\title{QCD Factorization of Semi-inclusive DIS process at Operator Level}

\author{Gao-Liang Zhou
}
\affiliation{Key Laboratory of Frontiers in
Theoretical Physics, \\The Institute of Theoretical Physics, Chinese
Academy of Sciences, Beijing 100190, People's Republic of China}




\begin{abstract}
The operator level proof of factorization theorem exhibited in \cite{zhougl} is extended to the semi-inclusive deep inelastic scattering process(SIDIS).
Factorization theorem can be proved at operator level if there are not
detected soft hadrons.
\end{abstract}

\maketitle

\section{Introduction.}

Factorization theorem plays important role in the understanding of high energy process involving hadrons.
The proof of this important theorem, however, turns out to be much nontrivial.
In conventional approach based on diagram level analyses, diagrams are decomposed in to two parts.
One of them takes the factorized form after the collinear and soft approximations are applied to it,
the other part is power suppressed. In \cite{zhougl}, we present an operator level proof
of factorization theorem for Drell-Yan process. Collinear fields that decouple from soft gluons are defined in \cite{zhougl} once the eikonal line approximations works
after the interactions  before the hard collision have been cancelled.
Effects of scalar-polarized collinear gluons are absorbed into Wilson lines in effective action that describe Hard process between different jets.
The factorization of soft and collinear particles can then be realized at the operator level.

In this paper, we extend the operator method to semi-inclusive deep inelastic process(SIDIS). It also makes the feature and procedure of this method more transparent. Such an operator method is closely related to the effective theory and is valuable for description of QCD factorization in the frame of effective field theory. We consider the collinear factorization in this paper, thus the back-to-back region do not disturb us.

Processes considered in this paper can be written as:
\begin{equation}
e+p\to e+H+X
\end{equation}
where $H$ denotes the detected collinear hadrons, $X$ denotes the undetected hadrons.
$H$ can be collinear to one initial hadron or moving in other directions.
The hadronic tensor for these processes reads:
\begin{eqnarray}
H^{\mu\nu}(q,H)
&=&\sum_{X}\int\ud^{4}x e^{-iq\cdot x}
\big<p|J^{\mu}(x)|HX\big>
\nonumber\\
&&
\big<HX|J^{\nu}(0)|p\big>
\end{eqnarray}
where $q^{\mu}$ is the transfer momentum of the electron.  We also define that $Q=\sqrt{|q^{2}|}$.

If there are not detected hadrons with small transverse momenta,
then the difference between semi-inclusive processes and inclusive processes is
that fragmentation functions should be brought in in the former case.
Collinear factorization of such processes at diagram level can be found in in \cite{CSS1989}.

If there are detected hadrons that with small transverse momenta, the situation is more complicated. It is well-known that QCD factorization is violated in the diffractive processes with initial hadrons moving in different directions. (\cite{CFS1993,CDF1997,ACTW1997,G1997,H12007}) This is because that cancellation of interactions after the collision as in \cite{B1985,CSS1985,CSS1988} is no longer the case for such processes. For processes with one initial hadron
and detected hadrons with small transverse momenta, for example the hard diffractive scattering in
high energy $ep$ collisions(\cite{H12007,diffractive}), factorization theorem is proved at diagram level
in \cite{C1998} and is confirmed by measurements of diffractive deep inelastic
scattering(\cite{H12007,diffractive}). Fracture functions\cite{TV1994} or diffractive parton distribution
functions\cite{BS1994} involve in such processes compared to those structure functions or parton distribution functions in inclusive processes.

The operator level proof of QCD factorization for such processes in this paper is organized as follows.
We first prove the cancellation of interactions before the collision in the hadronic tensor
in Sec.\ref{evolution}. The crucial point is  that the initial state is the eigenstate of the operator $\lim_{T\to\infty}e^{-iH_{QCD}2T}$. That is to say, one-hadron states at $t=-\infty$ will evolve to the same states at $t=\infty$ if there are not electro-weak interactions. This is indeed the case for nucleons as such states can not decay to other states if there are not electro-weak interactions.

After this cancellation, we consider the time evolution of the effective currants in Sec.\ref{deformation}. We show that one can always deform the integral path so that eikonal line works while calculating such evolution according to the similar method in \cite{zhougl}.

In Sec.\ref{effective theory}, we bring in the Wilson lines of scalar polarized gluons that absorb the effects of soft gluons. We also construct effective action that describe the electromagnetic scattering processes between different jets in this section.  The results are similar with those in \cite{zhougl}. The difference is that
the Wilson line of soft and collinear gluons travels from $x$ to $\infty$ in this case.

We consider the hadronic tensor in the frame of effective theory in Sec.\ref{factorization}.
Factorization theorem is then proved at the operator level.
\\

\section{Cancellation of Interactions Before the Hard Collision}
\label{evolution}

In this section, we will show the cancellation of interactions after the hard collision
in SIDIS.

In the following paragraphs, we divide particles into four classes.
The first class denoted by the subscript $d$ is the class of collinear particles with momenta
in the direction parallel or nearly parallel to the initial hadron.
The second class denoted by the subscript $n_{H}$ is the class of collinear particles with momenta
in the direction parallel or nearly parallel to one detected hadron and  quite different
from that of the initial hadron. The third class denoted by the subscript $\widetilde{n}$
is the class of collinear particles with momenta in the direction  different
from those of initial and detected hadrons.
The fourth class denoted by the subscript $s$ is the class of soft particles

We start from the hadronic tensor of SIDIS:
\begin{eqnarray}
H^{\mu\nu}(q,H)
&=&\sum_{X}\int\ud^{4}x e^{-iq\cdot x}\big<p|J^{\mu}(x)|HX\big>
\nonumber\\
&&
\big<HX|J^{\nu}(0)|p\big>
\end{eqnarray}
It can be dealt with according to a similar method in \cite{zhougl}, although there may be slightly differences for different processes. We write the hadronic tensor in Schr$\ddot{o}$inger picture:
\begin{eqnarray}
&&H^{\mu\nu}(q,H)
\nonumber\\
&=&\lim_{T\to\infty}\sum_{X} \int_{-T}^{T}\ud x^{0}\int\ud^{3}\vec{x} e^{-iq\cdot x}
\nonumber\\
&&
    _{-}\big<p|
    e^{iH_{QCD}(x^{0}+T)}J^{\mu}(\vec{x})e^{iH_{QCD}(T-x^{0})}|HX\big>_{+}
   \nonumber\\
&& _{+}\big<HX|e^{-iH_{QCD}T}J^{\nu}(\vec{0})e^{-iH_{QCD}T}
   |p\big>_{-}
\end{eqnarray}
where $+$ and $-$ denote the states at $t\to \pm \infty$. $|p\big>$ should be the eigenstate of the operator $e^{-iH_{QCD}2T}$. That is, if there are not the electro-weak interactions or other particles that interact with the initial hadron then the initial hadron with momentum $p$ at the time $t=-\infty$ should evolve to the same state(there can be different phase) at the time $t=\infty$. This is indeed the case if the initial hadron is a nucleon as it can not decay if there are not the electro-weak interactions.  We have:
\begin{eqnarray}
e^{-iH_{QCD}(x^{0}+T)}|p\big>_{-}&=&e^{iH_{QCD}(T-x^{0})}e^{-iH_{QCD}2T}|p\big>_{-}
\nonumber\\
&=&e^{-i\alpha(p,T)}e^{iH_{QCD}(T-x^{0})}|p\big>_{-}
\end{eqnarray}
with $\alpha$ the phase angle.
The hadronic tensor can be written as:
\begin{eqnarray}
&&H^{\mu\nu}(q,H)
\nonumber\\
&=&\lim_{T\to\infty}\sum_{X} \int_{-T}^{T}\ud x^{0}\int\ud^{3}\vec{x} e^{-iq\cdot x}
\nonumber\\
&&
    _{-}\big<p|
    e^{-iH_{QCD}(T-x^{0})}J^{\mu}(\vec{x})e^{iH_{QCD}(T-x^{0})}|HX\big>_{+}
   \nonumber\\
&& _{+}\big<HX|e^{-iH_{QCD}T}J^{\nu}(\vec{0})e^{iH_{QCD}T}
   |p\big>_{-}
\end{eqnarray}
We see that there are no longer interactions before the time $t=x^{0}\sim 1/Q$ or $t=0$
in this hadronic tensor. We then turn to the interaction picture and have:
\begin{eqnarray}
\label{Hmunu1}
H^{\mu\nu}(q,H)&=&
\sum_{n,n^{\prime},m, m^{\prime}}
\sum_{p_{i},p_{i}^{\prime},k_{i},k_{i}^{\prime},X}\int\ud^{4}x e^{-iq\cdot x}
\nonumber\\
&&
_{+}\big<p_{1}\ldots p_{n}|p\big>_{-}\quad
     _{-}\big<p|p_{1}^{\prime}\ldots p_{n^{\prime}}^{\prime}\big>_{+}
\nonumber\\
&&  _{+}\big<H|k_{1}\ldots k_{m}\big>_{+}
   \quad
   _{+}\big<k_{1}^{\prime}\ldots k_{m^{\prime}}^{\prime}|H\big>_{+}
   \nonumber\\
&&
    _{0} \big<p_{1}^{\prime}\ldots p_{n^{\prime}}^{\prime}|
    U_{QCD}(\infty,x^{0})J^{\mu}(x)
\nonumber\\
&&
    U_{QCD}^{\dag}(\infty,x^{0})|k_{1}^{\prime}\ldots k_{m^{\prime}}^{\prime}X\big>_{0}
   \nonumber\\
&& _{0}\big<k_{1}\ldots k_{m}X|U_{QCD}(\infty,0)J^{\nu}(0)
\nonumber\\
&&
U_{QCD}^{\dag}(\infty,0)|
  p_{1}\ldots p_{n}\big>_{0}
\end{eqnarray}
where the parton states $|X\big>_{0}$, $|p_{i}\big>_{0}$, $|p_{i}^{\prime}\big>_{0}$, $|k_{i}\big>_{0}$ and $|k_{i}^{\prime}\big>_{0}$ equal to corresponding  states in Schr$\ddot{o}$dinger picture at the time $t=0$, $|p_{1}\ldots p_{n}\big>_{0}=|p_{1}\big>_{0}\ldots |p_{n}\big>_{0}$, $U_{QCD}(t_{1},t_{2})$ is the time evolution operator of QCD in the interaction picture:
\begin{eqnarray}
U_{QCD}(t_{1},t_{2})&=&e^{iH_{0}t_{1}}e^{-iH_{QCD}(t_{1}-t_{2})}e^{-iH_{0}t_{2}}
\nonumber\\
                    &=&T\exp\{-i\int_{t_{2}}^{t_{1}}\ud t(H_{I})_{QCD}(t)\}
\end{eqnarray}
\\

\section{Deformation of Integral Path}
\label{deformation}
In this section we show that one can deform the integral path so that the eikonal line approximation works in (\ref{Hmunu1}). The conclusion is that while dealing with couplings between soft gluons and particles collinear to $n^{\mu}$, we can always deform the integral path of $n\cdot q$ to the lower half plane so that the eikonal line approximation works, where $q$ denote the momenta of Glauber gluons defined as flow into particles collinear to $n^{\mu}$.

In the hadronic tensor (\ref{Hmunu1}), $J^{\mu}(x)$ is local in time. Thus we can repeat the similar proofs as in \cite{zhougl} to show that ne can deform the integral path so that the eikonal line approximation works. We outline the procedure here:

(1) We first consider the evolution:
\begin{eqnarray}
\label{Mspace1}
&&U_{QCD}(\infty,x^{0})\psi(x)U_{QCD}^{\dag}(\infty,x^{0})\nonumber\\
&=&\psi(x)+\sum_{n}(-i)^{n}\int_{t_{n-1}}^{\infty}\ud t_{n}\ldots\int_{x^{0}}^{\infty}\ud t_{1}
\nonumber\\
&&[H_{I_{QCD}}(t_{n}),\ldots[H_{I_{QCD}}(t_{1}),\psi(x)]\ldots]
\nonumber\\
&=&\psi(x)+\sum_{n}\sum_{0\leq i<j\leq n}\sum_{G_{R},V_{n}^{i}}(-i)^{n}
\nonumber\\
&&
(\prod_{i=1}^{n}\int\frac{\ud^{4}q}{(2\pi)^{4}})
(\prod_{0\leq i\leq n-1}^{i<j\leq n}\int\frac{\ud^{4}k_{ij}}{(2\pi)^{4}})
\nonumber\\
&&e^{-i\sum_{i=1}^{n}q\cdot x}(\prod_{0\leq i\leq n-1}^{i<j\leq n}G_{R}(k_{ij}))
(\prod_{i=1}^{n}V_{n}^{i}(q_{i}))
\nonumber\\
&&
(\prod_{i=1}^{n}(2\pi)^{3}\delta^{(3)}(\vec{q}_{i}+\sum_{i<j\leq n}\vec{k}_{ij}-\sum_{0\leq j<i}\vec{k}_{ji}))
\nonumber\\
&&(\prod_{i=1}^{n}\frac{1}{-i(\sum_{j=i}^{n}q_{j}^{0}
-\sum_{i\leq j\leq n}^{0\leq l< i}k_{lj}^{0}-i\epsilon)})
\end{eqnarray}
where
\begin{equation}
G_{R}^{ij}(k)=
    \frac{i(\not\!k\pm m)^{ij}}{(k^{0}-E_{k}-i\epsilon)(k^{0}+E_{k}-i\epsilon)}
\end{equation}
for fermions and
\begin{eqnarray}
&&G_{R}^{\mu\nu}(k)
\nonumber\\
&=&
    \frac{-i}{(k^{0}-E_{k}-i\epsilon)(k^{0}+E_{k}-i\epsilon)}
\nonumber\\
&&
(g^{\mu\nu}-\frac{(1-\xi)k^{\mu}k^{\nu}}{(k^{0}-E_{k}-i\epsilon)(k^{0}+E_{k}-i\epsilon)})
\end{eqnarray}
for gluons in covariant gauge, $V_{n}^{i}$ are function of fermion fields and gluon fields that are not contracted with other fields. One should sum over all possible combination of $V_{n}^{i}$ and $G_{R}$, this is suggested by the third summation. We can also write the evolution as(\cite{zhougl}):
\begin{eqnarray}
\label{Mspace2}
&&U_{QCD}(\infty,x^{0})\psi(x)U_{QCD}^{\dag}(\infty,x^{0})\nonumber\\
&=&\psi(x)+\sum_{n}\sum_{0\leq i<j\leq n}\sum_{G_{R},V_{n}^{i}}C(G_{R},V_{n}^{i}) (-i)^{n}
\nonumber\\
&&
(\prod_{i=1}^{n}\int\frac{\ud^{4}q}{(2\pi)^{4}})
(\prod_{0\leq i\leq n-1}^{i<j\leq n}\int\frac{\ud^{4}k_{ij}}{(2\pi)^{4}})
\nonumber\\
&&e^{-i\sum_{i=1}^{n}k_{0i}\cdot x}(\prod_{0\leq i\leq n-1}^{i<j\leq n}G_{R}(k_{ij}))
(\prod_{i=1}^{n}V_{n}^{i}(q_{i}))
\nonumber\\
&&
(\prod_{i^{\prime}=1}^{n}(2\pi)^{3}\delta^{(3)}(\vec{q}_{i^{\prime}}+\sum_{i^{\prime}<j\leq n}\vec{k}_{i^{\prime}j}-\sum_{0\leq j<i^{\prime}}\vec{k}_{ji^{\prime}}))
\nonumber\\
&&
(\prod_{i^{\prime}=1}^{n}i
(q_{i^{\prime}}^{0}+\sum_{i^{\prime}<j\leq n}k_{i^{\prime}j}^{0}-\sum_{0\leq j<i^{\prime}}k_{ji^{\prime}}^{0}-i\epsilon)^{-1})
\end{eqnarray}
where $C(G_{R},V_{n}^{i})$ denote possible symmetrization factors which depends on the combination of $V_{n}^{i}$ and $G_{R}$.
For the evolution of $\bar{\psi}$, we have the similar result. we can then contract fields in $V_{n}^{i}$ terms with other fields in such terms or contract them with initial or final states to get the matrix-element.

(2)We then consider coupling between soft particles and collinear particles. If there are not collinear internal lines at the vertex, then one can simply drop the small momenta components in the collinear external and the collinear external lines is independent of momenta of soft particles.  We will assume that there are at least one collinear internal line at the vertex. We consider the case that soft gluons couple to particles collinear-to-plus at the point $x_{i}$. We notice that:
\begin{eqnarray}
\label{momenta coversation}
&&\frac{\delta(q_{i}^{3}+\sum_{i<i^{\prime}\leq n}k_{ii^{\prime}}^{3}-\sum_{0\leq i^{\prime}<i}k_{i^{\prime}i}^{3})}{q_{i}^{0}+\sum_{i<j^{\prime}\leq n}k_{ij^{\prime}}^{0}-\sum_{0\leq j^{\prime}<i}k_{j^{\prime}i}^{0}-i\epsilon}
\nonumber\\
&=&
\frac{1}{\sqrt{2}}\frac{\delta(q_{i}^{3}+\sum_{i<i^{\prime}\leq n}k_{ii^{\prime}}^{3}-\sum_{0\leq i^{\prime}<i}k_{i^{\prime}i}^{3})}{q_{i}^{-}+\sum_{i<j^{\prime}\leq n}k_{ij^{\prime}}^{-}-\sum_{0\leq j^{\prime}<i}k_{j^{\prime}i}^{-}-i\epsilon}
\end{eqnarray}
and make the approximation:
\begin{eqnarray}
\label{colliniear approximation}
&&\delta(q_{i}^{3}+\sum_{i<i^{\prime}\leq n}k_{ii^{\prime}}^{3}-\sum_{0\leq i^{\prime}<i}k_{i^{\prime}i}^{3})
\nonumber\\
&\simeq& \sqrt{2}\delta(\widetilde{q}_{i}^{+}+\sum_{i<i^{\prime}\leq n}\widetilde{k}_{ii^{\prime}}^{+}-\sum_{0\leq i^{\prime}<i}\widetilde{k}_{i^{\prime}i}^{+})
\end{eqnarray}
where $\widetilde{p}_{i}=p_{i}$ for collinear particles, $\widetilde{p}_{i}=(0,p_{i}^{-},(\vec{p}_{i})_{\perp})$ for soft particles.
After (\ref{momenta coversation}) and (\ref{colliniear approximation}), we see that singular points of $q_{i}^{-}$ and $k_{ij}^{-}$ in Glauber region that locate in the lower half plane can only be produced by the other end of $q_{i}^{+}$ and $k_{ij}^{-}$, while that of $k_{i^{\prime}i}^{-}$ that locate in the upper half plane can only be produced by the other end of $k_{i^{\prime}i}^{-}$.

(3)   If the other end of soft gluons connect to particles collinear to $n^{\mu}$ with $n^{3}=\cos(\theta)$, then singular point of $q_{s}$($q_{s}=q_{i}^{s}$, $k_{ij}$ or $k_{i^{\prime}i}$) produced by collinear internal lines at that end are those $n\cdot q_{s}\sim |(\vec{q}_{s})_{n\perp}^{2}|/Q$, where $\vec{n}_{\perp}$ denote the vector that fulfill the condition $\vec{n}_{\perp}\cdot\vec{n}=0$. We can then deform the integral path of $q_{i}^{s-}$ and $k_{ij}^{-}$ to lower half plane and that of $k_{i^{\prime}i}^{-}$ to upper half plane with radius of order:
 \begin{equation}
\min\{\frac{|(\vec{q}_{s})_{\perp}||\sin(\theta)|}{(1+\cos(\theta))},|\vec{q}_{s}|\}
\end{equation}.
After this deformation, we can drop the components $(\vec{q}_{s})_{\perp}$ in collinear internal lines at the point $x_{i}$ with corrections no greater than:
\begin{equation}
\max\{\frac{|(\vec{q}_{s})_{\perp}|(1+\cos(\theta))}{Q|\sin(\theta)|}\ast (1-\cos(\theta)),
\frac{|\vec{q}_{s}|}{Q}\}\lesssim \frac{|\vec{q}_{s}|}{Q}.
 \end{equation}
where we have assumed that $1-\xi$ is not too large with $\xi$ the gauge parameter, this is fulfilled in the Feynman gauge and Landau gauge.
We notice that $q_{i}$ and $k_{ij}$ are defined as flow in to the point $x_{i}$, $k_{i^{\prime}i}$ is defined as flow out of the point $x_{i}$, such deformation is in accordance with our claim at the beginning of this section.
\\

\section{Wilson Lines of Soft and Collinear Gluons}
\label{effective theory}

In this section we bring in Wilson-lines of scalar-polarized gluons to absorb effects of soft gluons and scalar-polarized collinear gluons.  They are defined according to the similar manner as in \cite{zhougl}, although the directions of Wilson lines involve in this paper is different from those in \cite{zhougl}. The action that describe hard sub-process  is also constructed in this section.

The hard collision is nearly local in coordinate space with uncertainty of order $1/Q$. As in \cite{zhougl},  we denote the space time region in which the hard collision occur as $H(x,1/Q)$, where $x$ is the point at which the hard photon interact with the target. We also denote the other space-time region as $S(x)$.   Different collinear jets separate from each other in the region $S(x)$.  We denote the part of $S(x)$ which jet collinear to $n^{\mu}$ locate in  as $x_{n}$.    We also denote the part of $S(x)$ in which there are not collinear jets as $x_{s}$. We consider the classical configuration of parton fields in SIDIS at this step. We define the soft fields as(\cite{zhougl}:
\begin{equation}
(D_{s\mu}G_{s}^{\mu\nu})^{a}(x_{n})=g\bar{\psi}_{s}\gamma^{\mu}t^{a}\psi_{s}(x_{n})
\end{equation}
\begin{equation}
 A_{s}^{\mu}(x_{s})=A^{\mu}(x_{s})
\end{equation}
and
\begin{equation}
\not\!D_{s}\psi_{s}(x_{n})=0,\quad \psi_{s}(x_{s})=\psi(x_{s})
\end{equation}
where
\begin{equation}
D_{s\mu}=\partial^{\mu}-ig A_{s}^{\mu},\quad G_{s}^{\mu\nu}=\frac{i}{g}[D_{s}^{\mu},D_{s}^{\nu}]
\end{equation}
where
\begin{equation}
D_{s\mu}=\partial^{\mu}-ig A_{s}^{\mu},\quad G_{s}^{\mu\nu}=\frac{i}{g}[D_{s}^{\mu},D_{s}^{\nu}]
\end{equation}
The equation should be solved perturbatively, that is, we define soft fields according to perturbation theory.
We also define the collinear fields as:
\begin{equation}
\psi_{n}(x_{n})=\psi(x_{n})
\end{equation}
\begin{equation}
\psi_{n}(x_{m})=\psi_{n}(x_{s})=0 (m^{\mu}\neq n^{\mu})
\end{equation}
\begin{equation}
A_{n}^{\mu}(x_{n})=A^{\mu}(x_{n})-A_{s}^{\mu}(x_{n})\end{equation}
\begin{equation}
 A_{n}^{\mu}(x_{m})=A_{n}^{\mu}(x_{m})=0 (m^{\mu}\neq n^{\mu})
\end{equation}

We then write the classical Lagrangian density of QCD as:
\begin{eqnarray}
\mathcal{L}_{QCD}&=&\sum_{n^{\mu}}\mathcal{L}_{n}(y)
+\mathcal{L}_{s}
\nonumber\\
&=&\sum_{n^{\mu}}i\bar{\psi}_{n}(\not\!\partial-ig\not\!A_{n}-ig\not\!A_{s})\psi_{n}
\nonumber\\
&&
-\frac{1}{2g^{2}}\sum_{n}^{\mu}tr_{c}\{[\partial^{\mu}-igA_{n}^{\mu}-igA_{s}^{\mu},
\nonumber\\
&&
\partial^{\nu}-ig A_{n}^{\nu}-ig A_{s}^{\nu}]^{2}\}
\nonumber\\
&&+i\bar{\psi}_{s}(\not\!\partial-ig\not\!A_{s})\psi_{s}
\nonumber\\
&&
-\frac{1}{2g^{2}}tr_{c}\{[\partial^{\mu}-i g A_{s}^{\mu},
\partial^{\nu}-igA_{s}^{\nu}]^{2}\})
\end{eqnarray}
We apply the eikonal line approximation to such Lagrangian density and have:
\begin{equation}
\partial^{\mu}-igA_{n}^{\mu}-igA_{s}^{\mu}
\simeq \widetilde{\partial}_{n}^{\phantom{n}\mu}-ig A_{n}^{\mu}-ign\cdot A_{s}\bar{n}^{\mu}
\end{equation}
where
\begin{equation}
\widetilde{\partial}_{n}^{\phantom{n}\mu}\psi_{n}=\partial^{\mu}\psi_{n}, \quad
\widetilde{\partial}_{n}^{\phantom{n}\mu}A_{n}^{\nu}=\partial^{\mu}A_{n}^{\nu}
\end{equation}
\begin{equation}
\widetilde{\partial}_{n}^{\phantom{n}\mu}n\cdot A_{s}=\bar{n}^{\mu}n\cdot\partial n\cdot A_{s}
\end{equation}
We then have:
\begin{equation}
\widetilde{\partial}_{n}^{\phantom{n}\mu}-ig A_{n}^{\mu}-ign\cdot A_{s}\bar{n}^{\mu}
=Y_{n}\widetilde{\partial}_{n}^{\phantom{n}\mu}Y_{n}^{\dag}-ig A_{n}^{\mu}
\end{equation}
where
\begin{equation}
\label{soft Wilson line}
Y_{n}(x_{n})=(P\exp(ig\int_{0}^{\infty}\ud s n\cdot A_{s}(x_{n}+sn))^{\dag}
\end{equation}
The Wilson line travel from $x_{n}$ to $\infty$, this is in accordance with our deformation of integral path of soft gluons. We redefine the fields:
\begin{equation}
\psi_{n}^{(0)}(x_{n})=Y_{n}^{\dag}\psi_{n}(x_{n})
\end{equation}
\begin{equation}
 A_{n}^{(0)\mu}(x_{n})=Y_{n}^{\dag}A_{n}^{\mu}(x_{n})Y_{n}(x_{n})
\end{equation}
and write $\mathcal{L}_{n}$ as:
\begin{eqnarray}
\mathcal{L}_{n}^{(0)}&=&i\bar{\psi}_{n}^{(0)}(\widetilde{\not\!\partial}_{n}
-ig\not\!A_{n}^{(0)})\psi_{n}^{(0)}(x_{n})
\nonumber\\
      &&+\frac{1}{2g^{2}}
      tr\left\{([\widetilde{\partial}_{n}^{\phantom{n}\mu}-igA_{n}^{(0)\mu},\right.
\nonumber\\
&&
        \left.\widetilde{\partial}_{n}^{\phantom{n}\nu}-igA_{n}^{(0)\nu}])^{2}\right\}(x_{n})
\end{eqnarray}
Thus $\psi_{n}^{(0)}$ and $A_{n}^{(0)}$ decouple from $A_{s}$ in $\mathcal{L}_{n}^{(0)}$. We denote the effective Lagrangian density in the region $S(x)$ as:
\begin{equation}
\mathcal{L}_{\Lambda}=\sum_{n^{\mu}}\mathcal{L}_{n}^{(0)}+\mathcal{L}_{s}
\end{equation}

$\mathcal{L}_{\Lambda}$ is invariant under the gauge transformation of $U_{s}(y)$ :
\begin{equation}
\psi_{s}(y)\to U_{s}\psi_{s}(y),\quad A_{s}^{\mu}(y)\to U_{s}(A_{s}^{\mu}+\frac{i}{g}\partial^{\mu})U_{s}^{\dag}(y)
\end{equation}
\begin{equation}
\psi_{n}^{(0)}(y)\to \psi_{n}^{(0)}(y),\quad
A_{n}^{(0)\mu}(y)\to A_{n}^{(0)\mu}(y)
\end{equation}
where we have constrained that $U_{s}(\infty)=1$.
We notice that the fields $\psi=\sum_{n\mu}\psi_{n}+\psi_{s}$ and $A=\sum_{n^{\mu}}A_{n}+A_{s}$ transform as:
\begin{equation}
\psi\to U_{s}\psi,\quad A^{\mu}\to U_{s}(A^{\mu}+\frac{i}{g}\partial^{\mu})U_{s}^{\dag}
\end{equation}
under $U_{s}$, thus this is just the usual gauge invariance of classical QCD Lagrangian density.
$\mathcal{L}_{\Lambda}$ is also invariant under the transformation:
\begin{equation}
\psi_{s}\to\psi_{s},\quad A_{s}^{\mu}\to A_{s}^{\mu}
\end{equation}
\begin{equation}
\psi_{n}^{(0)}\to U_{c}\psi_{n}^{(0)}
,\quad
A_{n}^{(0)\mu}\to U_{c}(A_{n}^{(0)\mu}+\frac{i}{g}\widetilde{\partial}_{n}^{\phantom{n}\mu})U_{c}^{\dag}
\end{equation}
This correspond to the gauge invariance of QCD in the special configuration $\psi_{s}=A_{s}=0$. There are not hadrons in the region $x_{s}$ in such special configuration.

We then quantize $\mathcal{L}_{\Lambda}$ by quantizing the effective fields $\psi_{n}^{(0)}$, $A_{n}^{(0)}$, $\psi_{s}$ and $A_{s}$ in stead of the usual parton fields $\psi$ and $A^{\mu}$. According to the similar arguments as in \cite{zhougl}, Such quantization scheme gives the same result as that in QCD at leading order in $M/Q$ while dealing with interactions between particles collinear to $n^{\mu}$ and soft particles.

We extend the effective fields to the region $H(x,1/Q)$, that is:
\begin{equation}
\psi_{n}^{(0)}(t,\vec{y})=e^{iH_{n}^{(0)}(t-t_{0})}\psi_{n}(t_{0},\vec{y})e^{-iH_{n}^{(0)}(t-t_{0})}
\end{equation}
\begin{equation}
A_{n}^{(0)(\mu)}(t,\vec{y})=
e^{iH_{n}^{(0)}(t-t_{0})}A_{n}^{(0)\mu}(t_{0},\vec{y})e^{-iH_{n}^{(0)}(t-t_{0})}
\end{equation}
\begin{equation}
\psi_{s}(t,\vec{y})=e^{iH_{s}(t-t_{0})}\psi_{s}(t_{0},\vec{y})e^{-iH_{s}(t-t_{0})}
\end{equation}
\begin{equation}
A_{s}^{(\mu)}(t,\vec{y})=e^{iH_{s}(t-t_{0})}A_{s}^{\mu}(t_{0},\vec{y})e^{-iH_{s}(t-t_{0})}
\end{equation}
where $(t,\vec{y})\in H(x,1/Q)$ and $(t_{0},\vec{y})\in S(x)$, $H_{n}^{(0)}$ and $H_{s}$ are the Hamiltonian corresponding to $\mathcal{L}_{n}^{(0)}$ and $\mathcal{L}_{s}$. To take the hard process into account, we bring in the effective operator $\Gamma^{\mu}(x)B_{\mu}(x)$ to describe such process, where $B^{\mu}$ denote the photon field. According to the gauge invariance under $U_{s}$, we have:
\begin{equation}
\Gamma^{\mu}(x)=\Gamma^{\mu}(Y_{n}\psi_{n}^{(0)}(x_{n}),\ldots,Y_{m}A_{m}^{(0)\mu}
Y_{m}^{\dag}(x_{m}))(x)
\end{equation}
where $x_{n}^{0}\geq x^{0}$, $x_{m}^{0}\geq x^{0}$ as there are not interactions before the production of hard photon in (\ref{Hmunu1}).  Couplings between soft particles and modes with $|k^{2}|\gtrsim Q^{2}$ are suppressed as power of $M/Q$, thus we can drop the $\psi(x_{s})$ and $A(x_{s})$ terms in $\Gamma^{\mu}$.

We work with the boundary condition $A(\infty)=0$ and write $\Gamma^{\mu}$ as
\begin{eqnarray}
\Gamma^{\mu}(x)
&=&\sum_{n^{\mu},\ldots, m^{\mu}}\int\ud (n\cdot x_{n})\int\ud (m\cdot x_{m})
\nonumber\\
&&
\mathcal{J}^{\mu}(\psi_{n}(x+n\cdot x_{n}\bar{n}),
\nonumber\\
&&
A_{m}^{\mu}
(x+m\cdot x_{m}\bar{m}))(x)
\end{eqnarray}
We have set $(x_{n})_{n\perp}=\bar{n}\cdot x_{n}=0$ as small momenta components of collinear particles can only contribute to the momenta conversation of the hard process in $\Gamma^{\mu}$.  The requirement $(x+x_{n})^{0}\geq x^{0}$  is equivalent to $n\cdot x_{n}\geq 0$ in this effective action as $\bar{n}\cdot x_{n}=0$.

To describe the effects of scalar-polarized gluons, we define the fields:
\begin{equation}
\widetilde{A}_{n,x_{0}}^{(0)\mu}(x_{n})
=A_{n}^{(0)\mu}(\bar{n}\cdot x_{0},n\cdot x_{n},(\vec{x}_{0})_{n\perp})
\end{equation}
\begin{equation}
\widetilde{A}_{n,x_{0}}^{\mu}(x_{n})
=A_{n}^{\mu}(\bar{n}\cdot x_{0},n\cdot x_{n},(\vec{x}_{0})_{n\perp})
\end{equation}
We also bring in the Wilson lines:
\begin{equation}
W_{n,x_{0}}^{(0)}(x_{n})=(P\exp(ig\int_{0}^{\infty}\ud s \bar{n}\cdot \widetilde{A}_{n,x_{0}}^{(0)}(x_{n}+s\bar{n})))^{\dag}
\end{equation}
\begin{equation}
W_{n,x_{0}}(x_{n})=(P\exp(ig\int_{0}^{\infty}\ud s
\bar{n}\cdot \widetilde{A}_{n,x_{0}}(x_{n}+s\bar{n}))^{\dag}
\end{equation}
They travel from $x_{0}$ to $\infty$, which are different from those in \cite{zhougl}. This is because that there are not interactions before the hard collision in the hadronic tensor (\ref{Hmunu1}).
For future convenience, we define the fields:
\begin{eqnarray}
\widehat{\psi}_{n,x_{0}}^{(0)}(x_{n})&=&W_{n,x_{0}}^{(0)\dag}(x_{n})\psi_{n}^{(0)}(x_{n})
\nonumber\\
(\partial^{\mu}-ig\widehat{A}_{n,x_{0}}^{(0)\mu})&=&
W_{n,x_{0}}^{(0)\dag}
(\partial^{\mu}-igA_{n}^{(0)\mu})
W_{n,x_{0}}^{(0)}
\end{eqnarray}
\begin{eqnarray}
\widehat{\psi}_{n,x_{0}}(x_{n})&=&W_{n,x_{0}}^{\dag}(x_{n})\psi_{n}(x_{n})
\nonumber\\
(\partial^{\mu}-ig\widehat{A}_{n}^{\mu})(x_{n})&=&
W_{n,x_{0}}^{\dag}
(\partial^{\mu}-igA_{n,x_{0}}^{\mu})
W_{n,x_{0}}
\end{eqnarray}
According to the similar method as in \cite{zhougl}, we write $\Gamma^{\mu}$ as:
\begin{eqnarray}
&&\Gamma^{\mu}|_{A(\infty)=0}
\nonumber\\
&=&\sum_{n^{\mu},\ldots, m^{\mu}}\ldots \int\ud (n\cdot x_{n})\int\ud (m\cdot x_{m})
\nonumber\\
&&
\mathcal{J}^{\mu}(\widehat{\psi}_{n,x}(x+n\cdot x_{n}\bar{n}),
\nonumber\\
&&\ldots,
(\partial^{m\perp}-\widehat{A}_{m,x}^{m\perp})(x+m\cdot x_{m}\bar{m}))
\end{eqnarray}
Physical fields in $\Gamma^{\mu}$ should connect to different jets at leading order.(\cite{zhougl,S1978}). Thus $\Gamma^{\mu}$ is multi-linear with its variables.

We then extract the large momenta components of collinear fields to make physical fields in $J^{\mu}$ be local in $x$. That is:
\begin{equation}
\psi_{n}(y)=\sum_{\bar{n}\cdot p}\psi_{n,\bar{n}\cdot p}(y)e^{-i\bar{n}\cdot p n\cdot y}
\end{equation}
\begin{equation}
A_{n}^{\mu}(y)=\sum_{\bar{n}\cdot p}A_{n,\bar{n}\cdot p}^{\mu}(x)e^{-i\bar{n}\cdot p n\cdot y}
\end{equation}
Then the large momenta components become labels on the effective fields. We take that $\psi_{n,\bar{n}\cdot p}(x+n\cdot x_{n}\bar{n})\simeq \psi_{n,\bar{n}\cdot p}(x)$ and $A_{n,\bar{n}\cdot p}^{\mu}(x+n\cdot x_{n}\bar{n})\simeq A_{n,\bar{n}\cdot p}(x)$ in $\Gamma^{\mu}$ as $n\cdot x_{n}\sim 1/Q$.  $\Gamma^{\mu}$ can then be written as:
\begin{eqnarray}
\Gamma^{\mu}|_{A(\infty)=0}
&=&
\sum_{n,\bar{n}\cdot p,\ldots,m,\bar{m}\cdot p^{\prime}}
\mathcal{J}^{\mu}((\widehat{\psi}_{n,x})_{\bar{n}\cdot p},\ldots,
\nonumber\\
&&
(\partial^{m\perp}-ig\widehat{A}_{n,x}^{m\perp})_{\bar{m}\cdot p^{\prime}}
)(x)
\nonumber\\
&=&
\sum_{n,\bar{n}\cdot p,\ldots,m,\bar{m}\cdot p^{\prime}}
\mathcal{J}^{\mu}(Y_{n}(\widehat{\psi}_{n,x}^{(0)})_{\bar{n}\cdot p},\ldots,
\nonumber\\
&&
Y_{m}(\partial^{m\perp}-ig\widehat{A}_{n,x}^{(0)m\perp})_{\bar{m}\cdot p^{\prime}}
Y_{m}^{\dag})(x)
\end{eqnarray}
If we consider the configuration  $\bar{n}\cdot \widetilde{A}_{n,x}=0$ (one should notice that $\bar{n}\cdot \widetilde{A}_{n,x}=0$ is not the axial gauge, it is just the lowest perturbation of $\bar{n}\cdot \widetilde{A}_{n,x}$), then $\mathcal{L}_{Q}$ can be matched perturbatively in the on-shell scheme.

We pause here to give some comments about the effective theory appeared in this paper. It seems similar with the Soft-Collinear Effective Theory (SCET) in \cite{SCET,SCET2}.  Especially, the Wilson lines and the fields redefinition  appeared these two methods are quite similar with each other. However, there are important differences between these two methods. In SCET, collinear modes and soft(ultra-soft) modes are distinguished according to their momenta. In this paper, we consider the classical configurations of parton fields before or after the hard collision. Fields in the space time region in which there are not collinear jets are defined as soft fields.  Thus momenta of the soft fields can be soft, ultrasoft and  Glauber at classical level. (That is to say, there can be soft, ultrasoft and Glauber gluons in soft hadrons
) Soft fields in the region in which there are collinear jets are defined according to perturbation theory. That is, soft gluons are defined as gluons exchanged between the region $x_{n}$ and $x_{s}$. The soft fields appeared in our method include the soft, ultrasoft and Glauber modes. Howler, while dealing with the hadronic tensor(\ref{Hmunu1}), one can take the eikonal line approximation to describe the coupling between soft particles and collinear particles as displayed in Sec.\ref{deformation}. We thus bring in the Wilson lines and redefinition of effective fields in the classical level. We then quantize the fields $\psi_{n}^{(0)}$, $A_{n}^{(0)}$, $\psi_{s}$ and $A_{s}$ in stead of the usual parton fields $\psi$ and $A^{\mu}$. This is different from SCET, as collinear fields are first quantized in the back ground gauge and then redefined by absorbing Wilson line of soft gluons. Thus, such two methods are different from each other even in the classical level.
\\

\section{Factorization.}
\label{factorization}

In this section, we finish the proof of QCD factorization for processes we are considering in this paper.

We write the hadronic tensor in the effective theory:
\begin{eqnarray}
\label{Hmunu}
&&H^{\mu\nu}(q,H)
\nonumber\\
&=&\lim_{T\to\infty}\sum_{X}\int\ud^{4}x e^{-iq\cdot x}
\nonumber\\
&&  _{-}\big<p|
    e^{-iH_{\Lambda}(T-x^{0})})\Gamma^{\mu\dag}(\vec{x})e^{iH_{\Lambda}(T-x^{0})})
    |HX\big>_{+}
   \nonumber\\
&& _{+}\big<HX|e^{-iH_{\Lambda}T}\Gamma^{\nu}(\vec{0})
e^{iH_{\Lambda}T}
   |p\big>_{-}
\end{eqnarray}
where we have made the directions $n^{\mu}$ slightly space-like, so that the time ordering and anti-time ordering operators do not affect the Wilson lines.

We write $\Gamma^{\mu}$ as:
\begin{eqnarray}
\Gamma^{\mu}(x)&=&\sum_{\Gamma_{c}^{\mu},\Gamma_{s}}^{color\quad indices}
\Gamma_{c}^{\mu}(\widehat{\psi}_{d,x}^{(0)},\widehat{\psi}_{n_{H}^{i},x }^{(0)},
\widehat{\psi}_{\widetilde{n}^{k},x }^{(0)},
\nonumber\\
&&
\partial^{n_{H}^{j}\perp}-ig\widehat{A}_{n_{H}^{j},x }^{(0)n_{H}^{j}\perp},
\partial^{\widetilde{n}^{l}\perp}
-ig\widehat{A}_{\widetilde{n}^{l},x}^{(0)\widetilde{n}^{l}\perp})
\nonumber\\
&&
\Gamma_{s}(Y_{d},Y_{n_{H}^{i} },Y_{n_{H}^{j}},Y_{\widetilde{n}^{k}},
Y_{\widetilde{n}^{l}})(x)
\end{eqnarray}
where $n_{H}^{i}$ and $\widetilde{n}^{k}$ represent the directions of $i$-th and $k$-th hadrons belong to the class $n_{H}$ and $\widetilde{n}$.
$\Gamma_{c}^{\mu}$ is multi-linear with $\widehat{\psi}_{n,x}^{(0)}$, $\widehat{\psi}_{m,x}^{(0)}$ and $\widehat{A}_{l,x }^{(0)\mu}$. Wilson lines that appear in $\Gamma_{s}$ depend on type of partons that appear in $\Gamma_{c}$. We notice that $(s_{1}n^{\mu}-s_{2}m^{\mu})^{2}<0$ for $s_{1}>0$, $s_{2}>0$ and $n^{\mu}\neq m^{\mu}$.  Thus, the order of Wilson lines in $\Gamma_{s}$ do not affect the result.

We consider the collinear factorization in this paper, thus we can set that $x^{\mu}=(n\cdot x,0,\vec{0})$ in the fields $\widehat{\psi}_{n,x}^{(0)}$ and $\widehat{A}_{n,x}^{(0)}$ and the $H_{n}x^{0}$ term. We can also set that $x=0$ in the Wilson lines $Y_{n}(x)$ and the $H_{s}x^{0}$ term.  The part of $H^{\mu\nu}$ that depend on soft fields are:
\begin{equation}
H^{s}(0)=\big<0|\Gamma_{s}^{\dag}(0)\Gamma_{s}(0)|0\big>
\end{equation}
According to unitarity of the Wilson lines $Y_{n}$ and colorlessness of hadrons, we have:
\begin{equation}
H^{s}(0)=1
\end{equation}
\begin{eqnarray}
&&H^{\mu\nu}(q,H)
\nonumber\\
&=&\lim_{T\to\infty}\sum_{X}\int\ud^{4}x e^{-iq\cdot x}
\nonumber\\
&&  _{-}\big<p|
    e^{-iH_{\Lambda}(T-x^{0})}\Gamma^{\mu\dag}(\vec{x})e^{iH_{\Lambda}(T-x^{0})})
    |HX\big>_{+}
   \nonumber\\
&& _{+}\big<HX|e^{-iH_{\Lambda}T}\Gamma^{\nu}(\vec{0})
e^{iH_{\Lambda}T}
   |p\big>_{-}|_{\psi_{s}=A_{s}=0}
\end{eqnarray}

We then define the annihilation operator(\cite{zhougl}):
\begin{eqnarray}
\label{fermion}
\widehat{a}_{n,x,p}^{s}&=&\frac{\sqrt{2E_{p}}}{2m}\int\ud^{3}\vec{x}_{n}
                             \bar{u}^{s}(p)\widehat{\psi}_{n,x}(\vec{x}_{n})
                             e^{-i\vec{p}\cdot \vec{x}}
\end{eqnarray}
\begin{eqnarray}
\label{antif}
\widehat{a}_{n,x,p}^{s}&=&-\frac{\sqrt{2E_{p}}}{2m}\int\ud^{3}\vec{x}_{n}
                   \bar{\widehat{\psi}}_{n,x}(\vec{x}_{n})
                   v^{s}(p)e^{-i\vec{p}\cdot \vec{x}}
\end{eqnarray}
\begin{eqnarray}
\label{gaugeb}
\widehat{a}_{n,p}^{j}
&=&\frac{i}{\bar{n}\cdot p}\sqrt{2E_{p}}
\int\ud^{3}\vec{x}_{n}e^{-i\vec{p}\cdot \vec{x}}
\nonumber\\
&&
\epsilon_{\mu}^{j*}(p)W_{n,x}^{\dag}G_{n}^{n\mu}
W_{n,x}(\vec{x}_{n})
\end{eqnarray}
with $j$ denotes different polarizations, where
\begin{equation}
G_{n}^{n\mu}=\frac{1}{-ig}
[\bar{n}\cdot(\partial-igA_{n}),
\partial^{\mu}-igA_{n}^{\mu}]
\end{equation}
States annihilated by these operators are denoted as $|\widehat{p}_{n}\big>_{x}$:
\begin{equation}
|\widehat{p}_{n}\big>_{x}=\sqrt{2E_{p_{n}}}\widehat{a}_{n,x,p_{n}}^{\dag}|0\big>
\end{equation}
where $\widehat{a}_{n,x,p_{n}}^{\dag}$ denote the conjugation of the operators (\ref{fermion}), (\ref{antif}) or (\ref{gaugeb}). We can then expand hadrons according to these states.

There can be no more than one parton states $|\widehat{p}_{n}\big>_{x}$ that contract with $\Gamma^{\mu}(x)$ for the jet collinear to $n^{\mu}$.
Without loss of generality, we denote these active partons as $|\widehat{p}^{1}\big>$ and  $|\widehat{k}_{i}^{1}\big>$
The hadronic tensor can then be written as:
\begin{eqnarray}
\label{facto}
&&H^{\mu\nu}(q,H)
\nonumber\\
&=&\sum_{\Gamma}
\sum_{Y}\sum_{p^{1},k_{i}^{1}}\frac{1}{N_{c}}\frac{1}{D_{G}}
\int\ud^{4}x e^{-iq\cdot x}
\nonumber\\
&&(2E_{p^{1}})tr_{c}\{\quad_{-}\big<p|e^{-iH_{\Lambda}^{d}(T-x^{0})}
        \widehat{a}_{p^{1}}^{\dag}e^{iH_{\Lambda}^{d}(T-x^{0})}|H_{d}Y_{d}\big>_{+}
\nonumber\\
&&
   _{+}\big<H_{d}Y_{d}|e^{-iH_{\Lambda}^{d}T}\widehat{a}_{p^{1}}
   e^{iH_{\Lambda}^{d}T}|p\big>_{-}\}|_{\psi_{s}=A_{s}=0}
   \nonumber\\
&& (\prod_{i}(2E_{k_{i}^{1}}) tr_{c}\{\big<0|
   \widehat{a}_{k_{i}^{1}}e^{iH_{\Lambda}^{n_{H}^{i}}(T-x^{0})}
   |H_{n_{H}^{i}}Y_{n_{H}^{i}}\big>_{+}
\nonumber\\
&&
 _{+}\big<H_{n_{H}^{i}}Y_{n_{H}^{i}}
   |e^{-iH_{\Lambda}^{n_{H}^{i}}T}\widehat{a}_{k_{i}^{1}}^{\dag}|0\big>\})|_{\psi_{s}=A_{s}=0}
\nonumber\\
&&
\nonumber\\
&& tr_{c}\{\big<\widehat{p}^{1}|\Gamma^{\mu\dag}(\vec{x})e^{iH_{\Lambda}^{\widetilde{n}}(T-x^{0})}
|Y_{\widetilde{n}}\ldots \widehat{k}_{i}^{1}\ldots\big>
\nonumber\\
&&
 \big<Y_{\widetilde{n}}\ldots \widehat{k}_{i}^{1}\ldots|e^{-iH_{\Lambda}^{\widetilde{n}}T}
   \Gamma^{\nu}(\vec{0})|\widehat{p}^{1}\big>\}|_{\psi_{s}=A_{s}=0}
\end{eqnarray}
where $H_{n_{H}^{i}}$ denotes detected hadrons collinear to the light-like direction $n_{H}^{i}$
with $n_{H}^{i}$ quite different from that of initial hadron, $Y$ denotes arbitrary states, $\frac{1}{D(G)}$ denotes the color factors produced by fields that do not collinear to initial hadron, $H_{\Lambda}^{d}$, $H_{\Lambda}^{n_{H}}$ and $H_{\Lambda}^{\widetilde{n}}$ denote the part of the $H_{\Lambda}$ that describe the jets belongs to the class $d$, $n_{H}$ and $\widetilde{n}$ respectively.

As in \cite{zhougl},
we take the lowest perturbation of the fields $\bar{n}\cdot \widetilde{A}_{n,x}$($n\notin \widetilde{n}$) in the matrix-elements between parton states in (\ref{facto}).  We then have:
\begin{eqnarray}
\label{dis}
&&H^{\mu\nu}(q,H)
\nonumber\\
&=&\sum_{\Gamma}
\sum_{Y}\sum_{p^{1},k_{i}^{1}}\frac{1}{N_{c}}\frac{1}{D_{G}}
\int\ud^{4}x e^{-iq\cdot x}
\nonumber\\
&&(2E_{p^{1}})tr_{c}\{\quad_{-}\big<p|e^{-iH_{\Lambda}^{d}(T-x^{0})}
        \widehat{a}_{p^{1}}^{\dag}e^{iH_{\Lambda}^{d}(T-x^{0})}|H_{d}Y_{d}\big>_{+}
\nonumber\\
&&
   _{+}\big<H_{d}Y_{d}|e^{-iH_{\Lambda}^{d}T}
   \widehat{a}_{p^{1}}e^{iH_{\Lambda}^{d}T}|p\big>_{-}\}|_{\psi_{s}=A_{s}=0}
\nonumber\\
&& (\prod_{i}(2E_{k_{i}^{1}}) tr_{c}\{\big<0|
   \widehat{a}_{k_{i}^{1}}e^{iH_{\Lambda}^{n_{H}^{i}}(T-x^{0})}
   |H_{n_{H}^{i}}Y_{n_{H}^{i}}\big>_{+}
\nonumber\\
&&
    _{+}\big<H_{n_{H}^{i}}Y_{n_{H}^{i}}
   |e^{-iH_{\Lambda}^{n_{H}^{i}}T}\widehat{a}_{k_{i}^{1}}^{\dag}|0\big>\})|_{\psi_{s}=A_{s}=0}
   \nonumber\\
&&
\nonumber\\
&& tr_{c}\{\big<p^{1}|\Gamma^{\mu\dag}(\vec{x})e^{iH_{\Lambda}^{\widetilde{n}}(T-x^{0})}
|Y_{\widetilde{n}}\ldots k_{i}^{1}\ldots\big>
\nonumber\\
&&
   \big<Y_{\widetilde{n}}\ldots k_{i}^{1}\ldots|e^{-iH_{\Lambda}^{\widetilde{n}}T}
   \Gamma^{\nu}(\vec{0})
\nonumber\\
&&
   |p^{1}\big>\}|_{
   \bar{n}_{d}\cdot \widetilde{A}_{d,x}^{(0)}=\bar{n}_{H}\cdot \widetilde{A}_{n_{H},x}^{(0)}=
   \psi_{s}=A_{s}=0}
\end{eqnarray}
where $|p_{i}^{1}\big>$ is the usual partons produced by the operator $a_{p_{i}}^{\dag}=\widehat{a}_{n_{i},x,p_{i}}^{\dag}|_{\bar{n}\cdot \widetilde{A}_{n_{i},x}=0}$. The condition $\bar{n}\cdot \widetilde{A}_{n,x}=0$ should be treated as the lowest perturbation of the fields $\bar{n}\cdot \widetilde{A}_{n,x}$ not the axial gauge.

The hadronic tensor (\ref{dis}) is our final result.
Soft gluons and scalar polarized gluons that collinear to initial or one detected final hadron  decouple from the matrix-elements of
$\Gamma$ between parton states in (\ref{dis}). Such matrix-elements can be calculated according to perturbation theory with
the loop momenta restricted not to collinear with the initial hadron or detected final hadrons. The momenta components components $n_{d}\cdot p^{1}$, $(p^{1})_{n_{d}\perp}$, $n_{H}^{i}\cdot k_{i}^{1}$ and $(k_{i}^{1})_{n_{H}^{i}\perp}$ can be dropped out of this matrix-element at leading order.
\\

\section{Conclusion.}

We have finished the proof of factorization theorem for semi-inclusive deep inelastic scattering process at operator level. Cancellation of interactions before the hard collision is realized by noticing that initial one-nucleon-states should be eigenstates of the time evolution operator: $\lim_{T\to\infty}e^{-iH_{QCD}(2T)}$. We have assumed that the photon couple to target hadron through point-like hard vertexes. For processes that can not described by such a hard vertex,
for example the resolved(hadron-like) photon processes(e.g.\cite{H12007}),what we have proved is the cancelation of interactions before the resolution of the photons not the hard collision.

After this cancelation, we can deform the integral path to avoid the Glauber region according to the similar proof as in \cite{zhougl}. In this proof, we require that the photon should be hard, that is $Q^{2}\gg M^{2}$ so that the scattering process is approximately local in time. We then define collinear fields that decouple from soft gluons after the deformation integral path. These collinear fields are just fields that annihilate or produce partons in the collinear jets equipped with future-pointing Wilson lines of soft gluons. Effects of hard sub-process  are absorbed into effective actions.

It is the gauge field tensor $G_{n}^{n\mu}$ not the gauge potential $A_{n}^{\mu}$ in our definition of annihilation operators that correspond to gluons fields in (\ref{gaugeb}).  Thus the gluons fragmentation functions in (\ref{dis}) is gauge invariant. For the parton distribution part in (\ref{dis}), there are definite hadrons $H$ in the final states. This is in accordance with the fracture functions\cite{TV1994} or diffractive parton distribution functions\cite{BS1994} structures.
\\

\section*{Acknowledgments}
The author thanks Y. Q. Chen for for helpful discussions and important suggestions about the article.
This work was supported by the National Nature Science Foundation of China under grants No.11275242.

\end{document}